\documentclass{KapProc}
\usepackage{graphics}

\let\footnote\savefootnote
\let\footnotetext\savefootnotetext 
\let\footnoterule\savefootnoterule 

\kluwerbib
\startauthorindex

\begin{document}

\newcommand{\iraf}{{\sc iraf}}
\newcommand{\irspec}{{\sc irspec}}
\newcommand{\caspir}{{\sc caspir}}
\newcommand{\class}{{\sc class}}
\newcommand{\Class}{{\sc Class}}
\newcommand{\iras}{{\sl IRAS}}
\newcommand{\brg}{Br$\gamma$}
\newcommand{\jv}{{V}}
\newcommand{\jr}{{R}}
\newcommand{\jj}{{J}}
\newcommand{\jh}{{H}}
\newcommand{\jk}{{K}}
\newcommand{\jl}{{L}}
\newcommand{\jn}{{N}}
\newcommand{\jhk}{{\jj \jh \jk}}
\newcommand{\jhkl}{{\jj \jh \jk \jl}}
\newcommand{\degr}{\hbox{$^\circ$}}
\newcommand{\arcmin}{\hbox{$^\prime$}}
\newcommand{\arcsec}{\hbox{$^{\prime\prime}$}}
\newcommand{\utw}{\smash{\rlap{\lower5pt\hbox{$\sim$}}}}
\newcommand{\udtw}{\smash{\rlap{\lower6pt\hbox{$\approx$}}}}
\newcommand{\fd}{\hbox{$.\!\!^{\rm d}$}}
\newcommand{\fh}{\hbox{$.\!\!^{\rm h}$}}
\newcommand{\fm}{\hbox{$.\!\!^{\rm m}$}}
\newcommand{\fs}{\hbox{$.\!\!^{\rm s}$}}
\newcommand{\fdg}{\hbox{$.\!\!^\circ$}}
\newcommand{\farcm}{\hbox{$.\mkern-4mu^\prime$}}
\newcommand{\farcs}{\hbox{$.\!\!^{\prime\prime}$}}
\newcommand{\fp}{\hbox{$.\!\!^{\scriptscriptstyle\rm p}$}}
\newcommand{\x}[1]{\hspace*{#1mm}}
\def\sun{\hbox{$\odot$}}

%------------ article title  ------------------->>

% If you use \\'s , please supply an alternate version of the title
% in square brackets, i.e., 
%\articletitle[Communism, Sparta, and Plato]
%{COMMUNISM, SPARTA,\\ and PLATO}

\articletitle[]{Properties of post-AGB stars with IRAS colors typical of planetary nebulae. }

%% supply a shorter version of the title for the running head:
\chaptitlerunninghead{Properties of post-AGB stars}

\author{Griet C. Van de Steene}
\affil{Royal Observatory of Belgium, Ringlaan 3, 1180 Brussels, Belgium}
\email{gsteene@oma.be}

\author{Peter A.M. van Hoof}
\affil{CITA, University of Toronto, 60 St. George Street, Toronto, ON M5S 3H8, Canada}
\email{vanhoof@cita.utoronto.ca}

\author{Peter R. Wood}
\affil{RSAA, Australian National University, Canberra ACT 0200, Australia}
\email{wood@mso.anu.edu.au}

%% short abstract
\begin{abstract}

We have investigated the infrared properties of  a sample of
objects with \iras\ colors typical of planetary nebulae.
The selected objects have not yet evolved to the
planetary nebula stage since they have no detectable radio continuum emission
and they are therefore likely post-AGB stars.

\end{abstract}

%------------ body of article ------------------->>

\section{Introduction}

A most intriguing challenge is to understand how Asymptotic Giant
Branch (AGB) stars transform their surrounding mass-loss shells in a
couple of thousand years into the variety of shapes and sizes observed
in Planetary Nebulae (PNe).  There are a number of theories currently being
investigated.  No consensus about the dominant physical process
responsible for the shaping of PNe has emerged so far, but there is
agreement that the shaping occurs during the early AGB-to-PN transition stage.

The details of the rate of evolution and the strength of the stellar
wind during the AGB-to-PN transition phase are essential ingredients
in understanding the shaping of the PN. However, both are very poorly known,
either theoretically or observationally.  To remedy this we have
started to study a sample of \iras-selected post-AGB candidates.  

\section{Sample Selection}

Objects were selected from the \iras\ catalogue with far infrared colors
typical of PNe. Apart from PNe, only post-AGB stars are typically
found in this part of the color-color diagram (van Hoof et al.\ \cite{vHoof97}).
Furthermore, we
selected objects that were not detected in the radio continuum above
a detection limit of 3 mJy (Van de Steene \& Pottasch \cite{VdSteene93}). 
Hence we may assume that they have not yet evolved to the
PN stage.

\section{Observations}

We obtained JHKL images of the candidates with \caspir\ on the 2.3-m
telescope at Siding Spring Observatory in Australia. 
To search for hydrogen emission, we obtained \brg\ spectra with \irspec\ 
on the NTT at ESO.  
The objects showing \brg\ in emission were re-observed in the
radio continuum with the Australia Telescope Compact Array.  

\section{Results}

The \iras\ counterparts were identified on the basis of their position
and near infrared colors. Accurate positions were obtained by using
the USNO catalog.  The weather was photometric and accurate JHKL
magnitudes were determined for the \iras\ counterparts.  By combining
the JHKL and \iras\ photometry, various color-color diagrams were
constructed.  In these diagrams our objects show evidence for a very
large range of extinction.  The observed trends in the infrared colors are consistent with
the expected evolution of the circumstellar shell (van Hoof et
al.\ \cite{vHoof97}).

Of 16 positively identified objects, 6 show
\brg\ in emission, 7 in absorption, and in 3 no clear \brg\
absorption or emission was visible.  The absorption lines are very
narrow in six objects, indicating a low surface gravity. The objects
showing \brg\ in emission have a strong underlying continuum,
unlike normal PNe.

The fact that our objects were mostly selected
from the region in the \iras\ color-color diagram where PNe
are typically found may explain our higher detection rate of emission line
objects compared to previous studies.

None of the objects showing \brg\ in emission were detected above
a detection limit of 0.6~mJy/beam at 6~cm and 0.7~mJy/beam at 3~cm,
while they should have been easily seen, if the radio emission was
optically thin and Case~B recombination was applicable. We
argue that the \brg\ emission is likely due to ionization in the
post-AGB wind, present before the star is hot enough to ionize the AGB
shell (Van de Steene et al. \cite{VdSteene00}).

In the near- and far-infrared color-color diagrams, no distinction 
can be made between objects showing \brg\  in
emission, \brg\ in absorption, or a flat spectrum. 
Whether the positions of the objects in the color-color diagrams
can be directly related to the temperature and core mass of the
central star needs further investigation, but doesn't seem likely
based on our data.

\section{Color-color diagrams }

\begin{figure}
\resizebox{!}{6cm}{\includegraphics{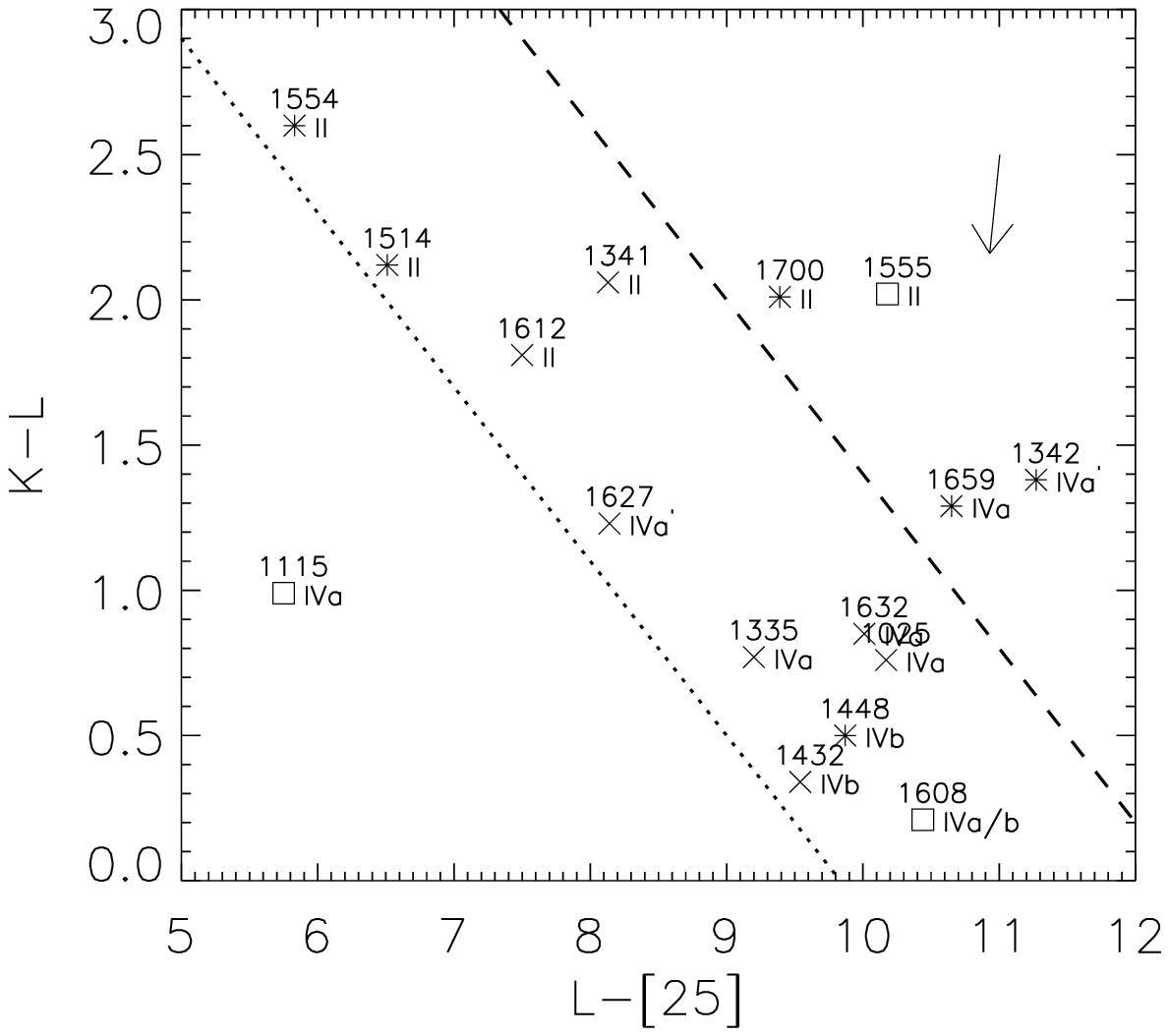} \includegraphics{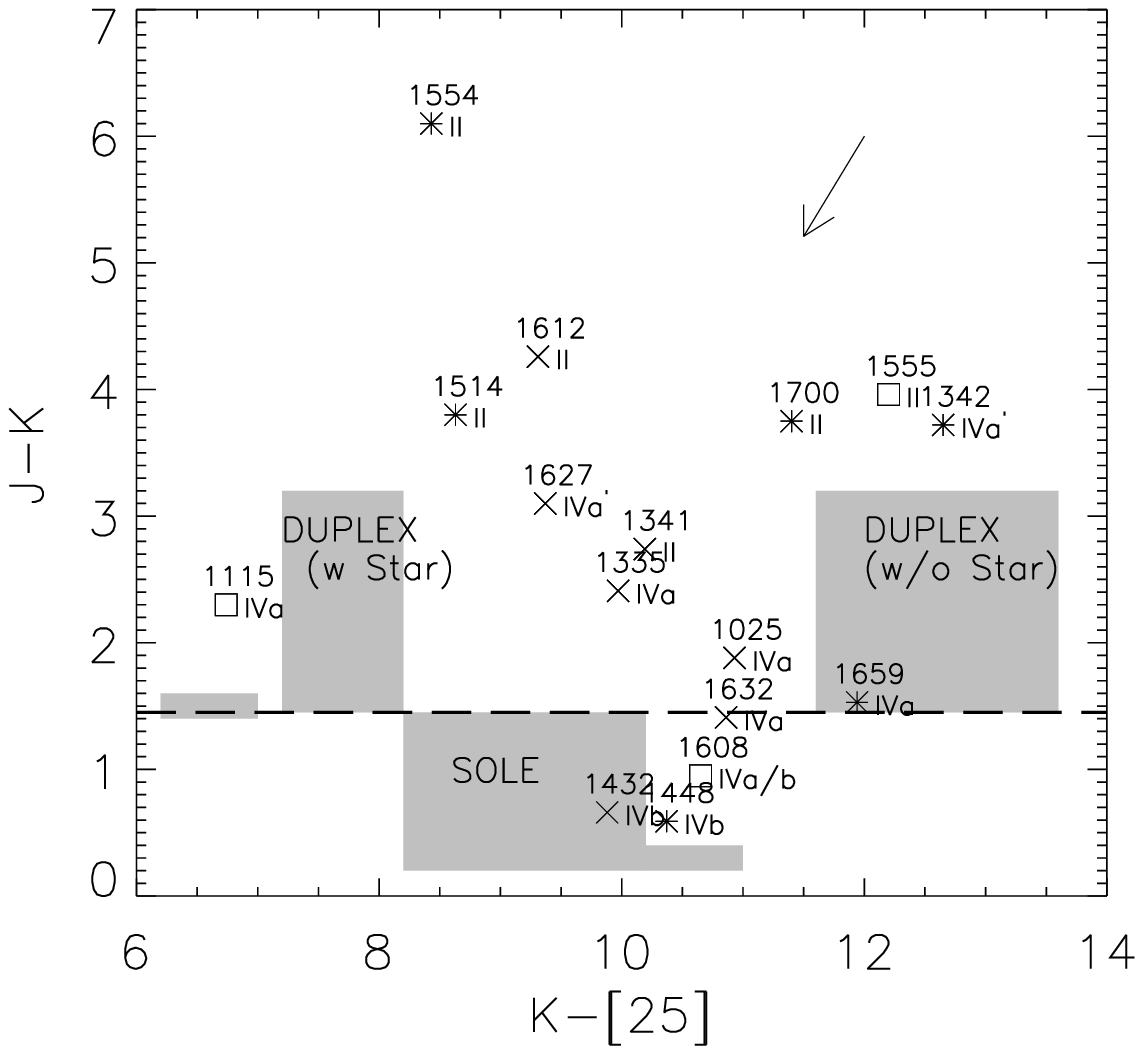}}
\dblcaption{The \jk-\jl,~\jl-[25] diagram. 
The crosses represent
objects having \brg\ in absorption, the asterisks objects having
\brg\ in emission, and the squares flat spectrum sources. The
objects are labeled by the first four numbers of their \iras\
name. The SED \class\ of each object is indicated next to its
symbol. The arrow shows the effect of reducing $A_V$ by 5~mag.  }
{ The \jj-\jk,~\jk-[25] diagram, as proposed by Ueta et al.\
(\cite{Ueta00}).  The grey regions show the location of the sample
of Ueta et al (\cite{Ueta00}).  These regions are labeled according to their
definitions as {\sc Star-Obvious Low level Elongated} ({\sc SOLE}) and
{\sc DUst Prominent Longitudinally EXtended} ({\sc DUPLEX}) with and
without star.  The dashed line separates these two kinds of objects.}
\end{figure}

In Figs.~1 and 2 we show two of the color
diagrams we have constructed.
The \jk$-$\jl\ and \jj$-$\jk\ colors follow the evolution of the
circumstellar extinction. The \jl$-$[25] and \jk$-$[25] colors relate
the stellar component with the peak of the dust emission, a larger color indicating
a more prominent dust shell. We note that
the \jl$-$[25] color is relatively insensitive to extinction.

In Fig.~1, stars obscured by their dust shells lie in the upper half
of the diagram (larger K$-$L) while stars visible through their dust shells lie
in the lower half.
The objects
located to the right of the dashed line are extended and some show
bipolar morphology in the K-band. Probably they have thicker, cooler
circumstellar dust shells than objects located below the dashed line
(more 25~$\mu$m-band flux) and/or their central star temperatures are
higher (less L-band flux).

For our limited sample, all but one of
the objects are found to the right of the dotted line.  This may reflect the evolution:
K$-$L decreases when mass loss ceases and the inner edge of the circumstellar shell expands,
and L becomes fainter as the central star becomes hotter at constant luminosity, causing
L$-$[25] to increase.

In Fig.~2 there is virtually no overlap between the sample of \\
Ueta et al.~(\cite{Ueta00}) and our sample.  This may be because of different
selection procedures for the two samples.  Ueta et al.~(\cite{Ueta00}) 
selected known post-AGB candidates from the
literature and imaged their nebulosities with WFPC2 in the optical.
We selected objects from the PNe region in the \iras\ color-color
diagram: very few of them have bright optical counterparts.

All our objects are located within 5 degrees of the galactic plane,
compared with only 11 of the 27 objects of  Ueta et al. (\cite{Ueta00}) (3 {\it SOLE}, 6 {\it
DUPLEX}, and 2 stellar).  The arrow in the diagram indicates an extinction
correction of $A_V$~= 5~mag.  Clearly, larger interstellar
extinction alone cannot explain why our samples appear different.
If they are at similar distances, it is plausible that we have more
massive central stars in our sample, and that they are evolving faster
across the HR diagram.
Further investigations will be needed to understand the differences
between both samples.

%------------ end of article ------------------->>

%% optional
% \begin{acknowledgments}
% text
% \end{acknowledgments}

%% remember to leave an empty line between bibliography imputs

\end{document}